\documentclass[a4paper,11pt]{article}
\pdfoutput=1 

\usepackage{jcappub} 

\usepackage{graphicx}
\usepackage{amsmath}
\usepackage{amssymb}
\usepackage{rotating} 
\usepackage{xspace}
\usepackage{color}
\usepackage{booktabs}
\usepackage[normalem]{ulem}

\newcommand{\Msun}{{\rm M}_\odot}

\title{Velocity-dependent annihilation radiation from dark matter subhalos in cosmological simulations}

\author[a]{Erin Piccirillo,}
\author[b]{Keagan Blanchette,}
\author[c, b]{Nassim Bozorgnia,}
\author[a]{Louis E. Strigari,}
\author[d]{Carlos S. Frenk,}
\author[e,f]{Robert J. J. Grand,}
\author[g]{and Federico Marinacci}

\affiliation[a]{Department of Physics and Astronomy, \\
Mitchell Institute for Fundamental Physics and Astronomy, \\
Texas A$\&$M University, College Station, TX 77843, USA}
\affiliation[b]{York University, Department of Physics and Astronomy,\\
4700 Keele Street, Toronto, Ontario M3J 1P3, Canada}
\affiliation[c]{Department of Physics, University of Alberta, \\
Edmonton, Alberta T6G 2E1, Canada}
\affiliation[d]{Institute for Computational Cosmology, Durham University,\\
South Road, Durham DH1 3LE, UK}
\affiliation[e]{Instituto de Astrof\'isica de Canarias, \\
Calle Vía L\'actea s/n, E-38205 La Laguna, Tenerife, Spain}
\affiliation[f]{Departamento de Astrof\'isica, Universidad de La Laguna, \\
Av. del Astrof\'isico Francisco S\'anchez s/n, E-38206, La Laguna, Tenerife, Spain}
\affiliation[g]{Department of Physics and Astronomy ``Augusto Righi'',\\ 
University of Bologna, via Gobetti 93/2, 40129 Bologna, Italy}

\abstract{We use the suite of Milky Way-like galaxies in the Auriga simulations to determine the contribution to annihilation radiation from dark matter subhalos in three velocity-dependent dark matter annihilation models: Sommerfeld, p-wave, and d-wave models. We compare these to the corresponding distribution in the velocity-independent s-wave annihilation model. For both the hydrodynamical and dark-matter-only simulations, only in the case of the Sommerfeld-enhanced annihilation does the total annihilation flux from subhalos exceed the total annihilation flux from the smooth halo component within the virial radius of the halo. Progressing from Sommerfeld to the s, p, and d-wave models, the contribution from the smooth component of the halo becomes more dominant, implying that for the p-wave and d-wave models the smooth component is by far the dominant contribution to the radiation. Comparing to the Galactic center excess observed by Fermi-LAT, for all simulated halos the emission is dominated by the smooth halo contribution. However, it is possible that for Sommerfeld models, extrapolation down to mass scales below the current resolution limit of the simulation would imply a non-negligible contribution to the gamma-ray emission from the Galactic Center region. 

}

\begin{document}
\maketitle
\flushbottom

\section{Introduction}
\label{sec:intro}
Dark matter (DM) particles may be detected via self-annihilation into gamma-rays within astrophysical environments~\cite{Conrad:2015bsa}. The gamma-ray emission from DM annihilation depends on the strength of the annihilation cross section, $\sigma_A$, and the phase-space distribution of DM within the system. The astrophysical dependence of the annihilation rate is encapsulated in a quantity typically denoted in the literature as the \emph{${\cal J}$-factor}, which is an integral over the dark matter phase-space distribution. For annihilation cross sections that are independent of relative velocity, or s-wave models, the \emph{${\cal J}$-factor} is simply reduced to an integral over the density squared of the dark matter distribution. If the annihilation cross section is velocity dependent, as in the cases of p-wave, d-wave, or Sommerfeld models, the ${\cal J}$-factor must account for this velocity dependence by incorporating the full dark matter velocity distribution~\cite{MarchRussell:2008tu, MarchRussell:2008yu,Robertson:2009bh,Ferrer:2013cla,Boddy:2017vpe,Zhao:2017dln,Petac:2018gue,Boddy:2018ike,Lacroix:2018qqh,Boddy:2019wfg,McKeown_2022, Hisano:2011dc}. 

\par The DM velocity distribution of a system depends on its dynamical properties, such as the potential, the formation history, and the possible non-equilibrium nature. Assuming the system is in dynamical equilibrium, a simple way to estimate the DM velocity distribution from the gravitational potential is to start from the DM density distribution, assume the system is isotropic, and use an Eddington inversion to obtain the velocity distribution~\cite{Evans:2005tn}. This can then be converted to the DM relative velocity distribution~\cite{Ferrer:2013cla}, which is an additional relevant quantity for indirect DM searches using gamma rays in the general cases of velocity-dependent DM annihilation. 

\par Simplified Eddington-type inversions, however, only determine the DM velocity distribution that is due to the smooth component of the DM potential. Therefore, they are not sensitive to contributions from DM substructures in the halo. Due to the large abundance of subhalos and their high central DM densities, DM subhalos are predicted to contribute to the gamma-ray emission from DM annihilation within the Milky Way (MW). This emission is in addition to that from luminous dwarf galaxies~\cite{Strigari:2018utn}, which have been used to set the most stringent and robust bounds on the DM annihilation cross section~\cite{Ackermann:2015zua,Boddy:2019wfg}.

\par Cosmological simulations provide the most direct method for estimating the DM velocity distribution, including the effects from DM substructures. Using the DM-only (DMO) Aquarius simulation, ref.~\cite{Springel:2008zz} showed that the gamma-ray emission from DM annihilation in subhalos is subdominant as compared to the emission from the smooth DM component in the MW. Using the suite of Auriga simulations~\cite{Grand:2016mgo}, which include the full impact of baryonic physics and accounting for the extrapolation of the halo concentration-mass relation to the smallest subhalo scales~\cite{Wang:2020}, ref.~\cite{Grand:2020} showed that the substructure contribution is even further reduced in simulations with baryons. This is likely due to baryonic effects facilitating the destruction of subhalos that orbit near the disk, as well as baryonic contraction producing a more dominant background. This has important implications for the x-ray~\cite{Zhong_2020} and gamma-ray emissions observed by the Fermi-LAT near the inner Galaxy~\cite{Grand:2022olu}. These results from the Aquarius and Auriga simulations are relevant for velocity-independent, s-wave DM annihilation models. 

\par Relative to s-wave models, it is more complicated to estimate gamma-ray emission from DM annihilation for the cases of velocity-dependent Sommerfeld, p-wave, and d-wave models, as this requires a direct estimate of the relative velocity distribution from the simulations. Using simulations of MW-like galaxies in the Auriga and APOSTLE suite of simulations, recent analysis of the smooth component of the DM distribution (excluding DM substructure) has shown that the relative DM velocity distribution is very well approximated by a Maxwell-Boltzmann distribution, with a velocity dispersion related to the circular speed at a given radius~\cite{Board:2021bwj}. This result then implies that for the smooth DM component of the halo, the \emph{${\cal J}$-factor} depends primarily on the DM density distribution, for different models of velocity-dependent DM annihilation. This has interesting implications for the phenomenology of DM signals for velocity-dependent annihilation models using Fermi-LAT~\cite{Smyth:2021bcp}. 

In this work, we extend the use of the pair-wise relative velocity distribution to DM subhalos and calculate a quantity analogous to the \emph{${\cal J}$-factor}, which we refer to as the annihilation luminosity. For our study, we use the simulated MW-like halos in the Auriga hydrodynamical simulations as well as their DMO counterparts. We calculate the DM relative velocity distribution for subhalos and use this distribution to determine the velocity-dependent annihilation lumonosities for the p-wave, d-wave, and Sommerfeld annihilation cross-section models.

The paper is organized as follows. In section~\ref{sec:simulations} we discuss the simulations that we analyze. In section~\ref{sec:properties} we determine the density profiles, maximum circular velocities, and relative velocity distributions for resolved DM subhalos from our simulations. In section~\ref{sec:luminosities} we introduce the formalism for computing the annihilation luminosities for the different DM annihilation models we consider. In section~\ref{sec:results} we present our results for the DM annihilation luminosity from the smooth halo, resolved DM subhalos, and extrapolated subhalos for each annihilation model. Finally, in section~\ref{sec:conclusion} we discuss the implications of our results and summarize our conclusions.

\section{Simulations}
\label{sec:simulations}
In this work we use hydrodynamical simulations of MW-like halos from the Auriga~\citep{Grand:2016mgo} project. 
The Auriga simulations~\citep{Grand:2016mgo} include a suite of thirty magneto-hydrodynamical zoom simulations of isolated MW mass halos, selected from a $100^3$~Mpc$^3$ periodic cube (L100N1504) from the EAGLE project~\cite{Schaye2015,Crain2015}. The simulations were performed using the moving-mesh code Arepo~\citep{Springel2010} and a galaxy formation subgrid model which includes star formation, feedback from supernovae and active galactic nuclei, metal-line cooling, and background UV/X-ray photoionisation radiation~\cite{Grand:2016mgo}. The cosmological parameters used for the simulations are from Planck-2015~\citep{Planck2015} measurements: $\Omega_{m}=0.307$, $\Omega_b=0.048$, $H_0=67.77~{\rm km~s^{-1}~Mpc^{-1}}$. In this work we use the high resolution level (Level 3) of the simulations with DM particle mass, $m_{\rm DM}=5 \times 10^4~\Msun$, baryonic mass, $m_b=6\times10^3~\Msun$, and Plummer equivalent gravitational softening length of $\epsilon =184$~pc~\citep{Power:2002sw,Jenkins2013}. The virial radius, $r_{200}$, is the radius that encloses a mean density 200 times the critical density and the virial mass is the total mass, $M_{200}$, enclosed within $r_{200}$. The virial mass, virial radius, and stellar mass of the Auriga halos are listed in table~\ref{tab:MWlike}.

All simulated halos have a DMO counterpart which share the same initial conditions as the hydrodynamical runs, but galaxy formation processes are ignored and the particles are treated as collisionless. We shall refer to halos in the hydrodynamical simulations as Auriga halos and those in the DMO simulations as DMO halos.

\begin{figure}[t]
    \centering
    \includegraphics[width=0.85\textwidth]{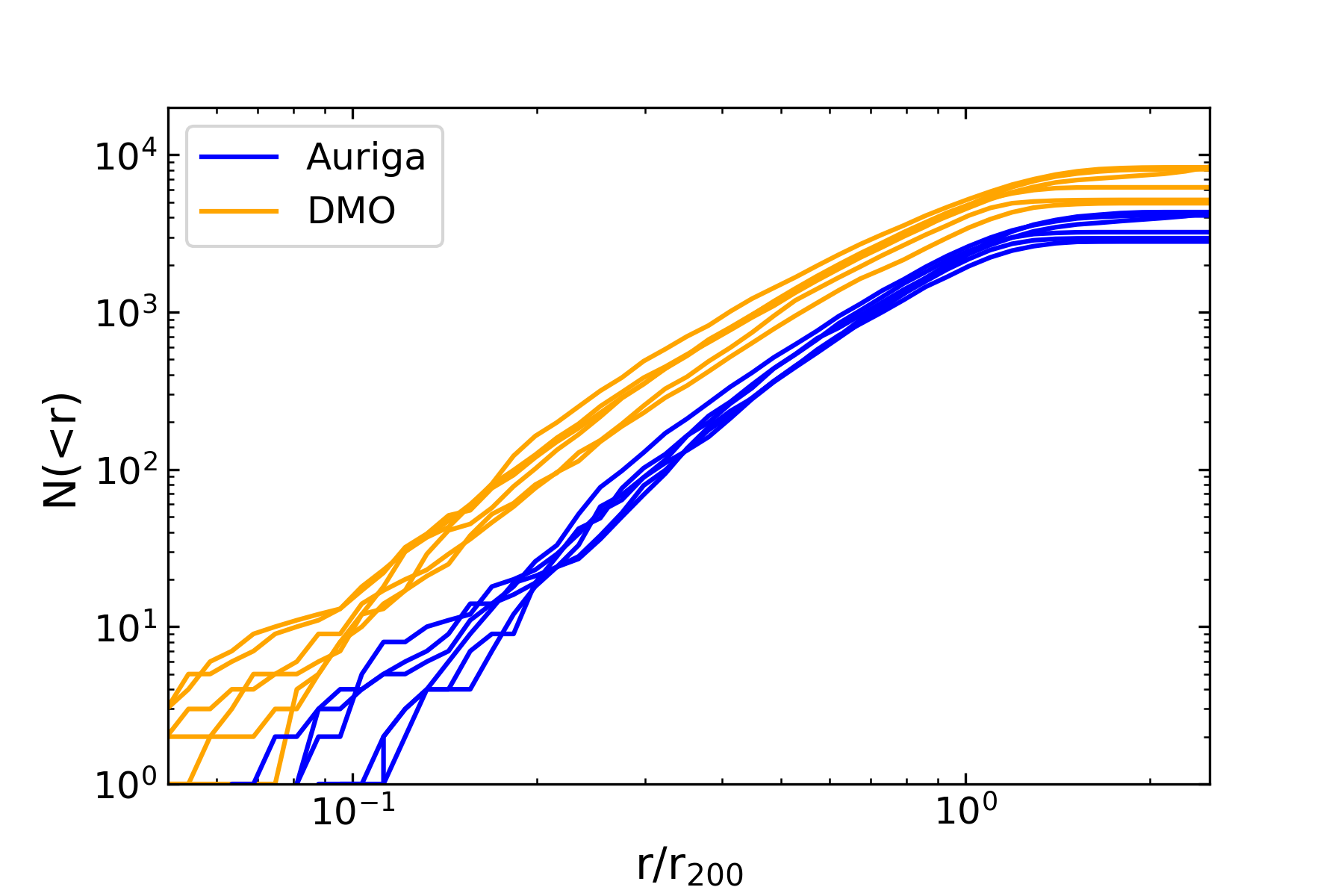}
    \caption{The cumulative number of subhalos enclosed within a sphere of Galactocentric radius $r$ for each Auriga halo (blue) and its DMO counterpart (orange). See table~\ref{tab:MWlike} for the virial radius, $r_{200}$, of each Auriga halo.}
    \label{fig:cumulativenumberofsubhalos}
\end{figure}

For the analysis in this work, we will consider DM particles bound to the smooth halo component as well as DM particles bound to subhalos, as identified by the SUBFIND algorithm~\cite{Springel:2000qu}. These simulations can resolve subhalos of mass greater than $\sim10^{6}~\Msun$, which contain at least 20 DM particles. Each Auriga halo contains $[1,925 - 2,405]$ DM subhalos within $r_{200}$, while each DMO halo contains $[3,377 - 5,157]$ DM subhalos within $r_{200}$. Figure~\ref{fig:cumulativenumberofsubhalos} shows the cumulative number of subhalos enclosed within a sphere of Galactocentric radius $r$ for each Auriga and DMO halo. We note that these results are dependent on the simulation used, and different baryonic effects such as a more prominent Milky Way disk can alter the number of subhalos in the simulation~\cite{Kelley:2018pdy}.

 \begin{table}[h!]
    \centering
    \begin{tabular}{|c|c|c|c|}
      \hline
      Halo Name  & $M_{\rm 200}~[\times 10^{12} \, \Msun]$ & $r_{200}~$[kpc] &  $M_\star~[\times 10^{10} \, \Msun]$ \\
      \hline
      Au6 & 1.01 & 211.8 & 6.08 \\
      Au16 & 1.05 & 241.5 & 7.85 \\
      Au21 & 1.50 & 236.7 & 8.19  \\
      Au23 & 1.49  & 241.5 & 8.28 \\
      Au24 & 1.50 & 239.6 & 7.77 \\
      Au27 & 1.47 & 251.4 & 9.47 \\
      \hline
    \end{tabular}
\caption{The virial masses, virial radii, and stellar masses (within $0.1 \times r_{200}$) of the Auriga MW-like halos, labeled by ``Au-Halo Number''.}
    \label{tab:MWlike}
  \end{table} 

\section{Properties of dark matter subhalos}
\label{sec:properties}
\par In this section, we discuss the dynamical properties of the DM subhalos that are most important for our analysis. We focus specifically on the DM density profiles, the maximum circular velocities, and the relative DM velocity distributions. These properties are then used in the subsequent sections to calculate the DM annihilation luminosity from each subhalo. 

\subsection{Density profiles}
\label{sec:Density profiles}

\begin{figure}[t]
    \centering
    \includegraphics[width=0.85\textwidth]{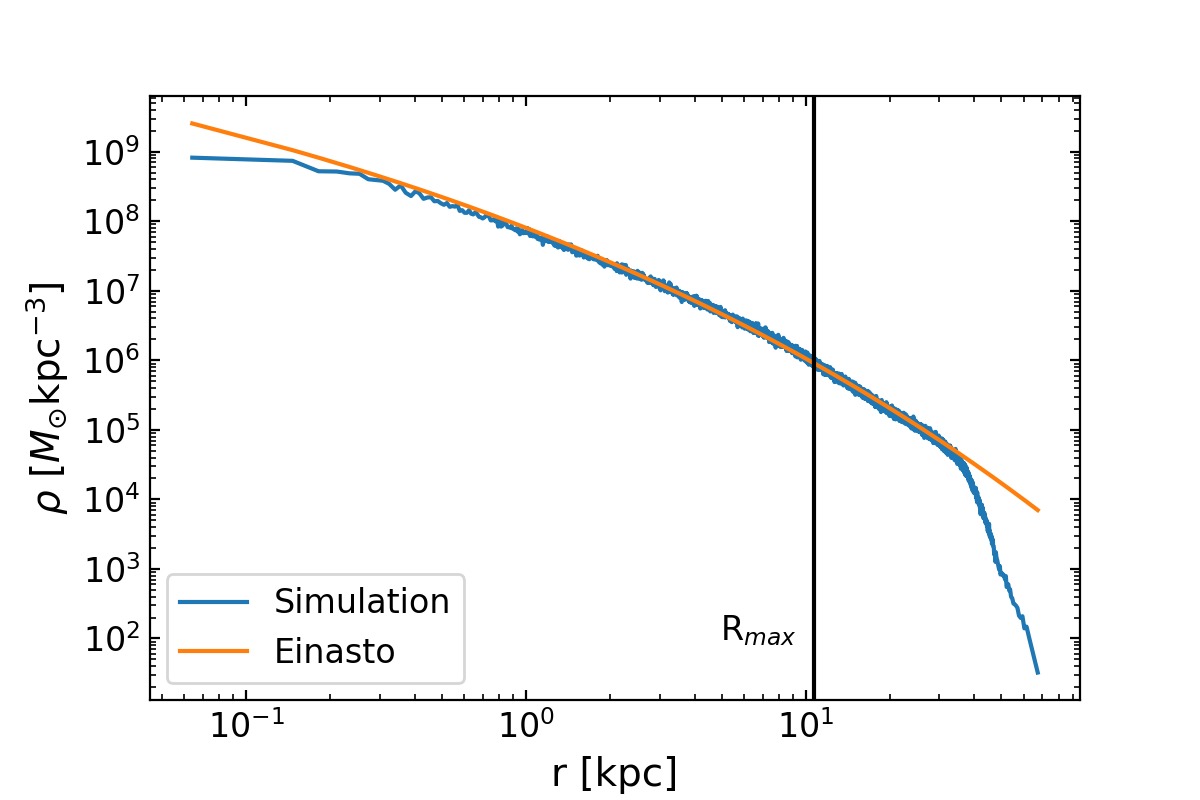}
    \caption{The DM density profile for a large subhalo in the Au6 simulation where $r$ is the distance from the center of the subhalo. The results from the simulation data and the best-fit Einasto profile are shown in blue and orange, respectively. This subhalo has a stellar mass of $4\times10^{9}~\Msun$, DM mass of $4\times10^{10}~\Msun$, and $V_{\rm max} = 75$ km/${\rm s}$ at $R_{\rm max} = 10.7$~kpc, which is indicated by the vertical black line. For large subhalos in the sky maps  in section~\ref{sec:luminosities}, we use the local DM density as estimated by a Voronoi tessellation for distances larger than $R_{\rm max}$ and use the best-fit Einasto density profile to estimate the local DM density for distances within $R_{\rm max}$.}
    \label{fig:subhalodensityprofile}
\end{figure}

\par For each subhalo, we obtain the spherically-averaged DM density profile from the DM mass contained within spherical shells centered on the center of potential of each subhalo as determined by SUBFIND. The number of DM particles per subhalo varies greatly between subhalos, with the minimum number being 20 DM particles bound to a subhalo. Thus, we use variable bin widths, ensuring that there is a minimum of 5 DM particles (for smaller subhalos) and a maximum of 200 DM particles (for larger subhalos) within each shell. For reasons discussed in section~\ref{sec:luminosities}, we define large subhalos to have an angular size $>1$ degree as viewed from the solar position, and small subhalos to have an angular size $<1$ degree. We fit the DM density profile constructed from the data to an Einasto density profile
\begin{equation}
\rho = \rho_{-2}  \exp\left(-\frac{2}{\alpha}\left[\left(\frac{r}{r_{-2}}\right)^\alpha - 1\right]\right),
\label{eq:Einasto density profile}
\end{equation}
where $\rho_{-2}$ and $r_{-2}$ are the density and radius at which $\rho(r) \propto r^{-2}$, and $\alpha$ is a parameter which specifies the curvature of the density profile. We set this parameter to $\alpha = 0.16$~\cite{Gao:2007gh}. In order to account for numerical resolution, we fit the Einasto density profile to the simulation data for radii larger than $2\epsilon$, where $\epsilon$ is the softening length defined in section~\ref{sec:simulations}. At large radii, we fit the profile up to a maximum radius of $2 R_{\rm max}^{\rm SUBFIND}$, where $R_{\rm max}^{\rm SUBFIND}$ is the radius of maximum circular velocity derived from the SUBFIND algorithm~\cite{Springel:2000qu}. Beyond $2 R_{\rm max}^{\rm SUBFIND}$, a large fraction of subhalos have their density profiles tidally stripped, such that they fall off faster than an Einasto density profile.

We also calculate $R_{\rm max}$ and $V_{\rm max}$ for each subhalo from the particle distribution. For an individual DM subhalo, most of the annihilation signal comes from within $R_{\rm max}$, the radius where the circular velocity $V_{c}(r)$ reaches a maximum, $V_{\rm max}$. For each subhalo, we calculate the circular velocity curve $V_{c}(r) = \sqrt{GM(<r)/r}$, where $M(<r)$ is the total DM mass enclosed within a sphere of radius $r$ centered on the subhalo. We find that our calculations of $V_{\rm max}$ and $R_{\rm max}$ are consistent with the values returned by SUBFIND. For internal consistency, we will use our calculations of $V_{\rm max}$ and $R_{\rm max}$ in this work. The $V_{\rm max}$ and $R_{\rm max}$ are used in section~\ref{sec:luminosities} to estimate the total annihilation luminosity within $R_{\rm max}$ for each subhalo.

\par Figure~\ref{fig:subhalodensityprofile} shows the DM density profile of one example subhalo from Auriga halo Au6, along with the best fit Einasto density profile for that subhalo. Also shown is the $R_{\rm max}$ value for the same subhalo. Due to the resolution limit of the simulation, the density profile calculated from the particle data underestimates the density in the central regions of the subhalos. For small subhalos in the sky maps  in section~\ref{sec:luminosities}, we will estimate the total DM annihilation luminosity using the calculated values of $V_{\rm max}$ and $R_{\rm max}$. For large subhalos in the sky maps, we use the best fit Einasto density profile for particles within $R_{\rm max}$ and we use the local DM density estimated by a Voronoi tessellation of the DM particle distribution for particles beyond $R_{\rm max}$. Following ref.~\cite{Grand:2020}, we apply a Voronoi tesselator to estimate the DM distribution in the outer radii of each subhalo, allowing the calculation of $\rho_{i}$ from the DM particle mass and the cell volume surrounding the $i$-th DM particle. For these large radii, this approach provides a better localized measure of the DM density than other estimates which smooth over a particle's nearest neighbors~\cite{grand2022dark}.

\begin{figure}[t]
    \centering
    \includegraphics[width=0.85\textwidth]{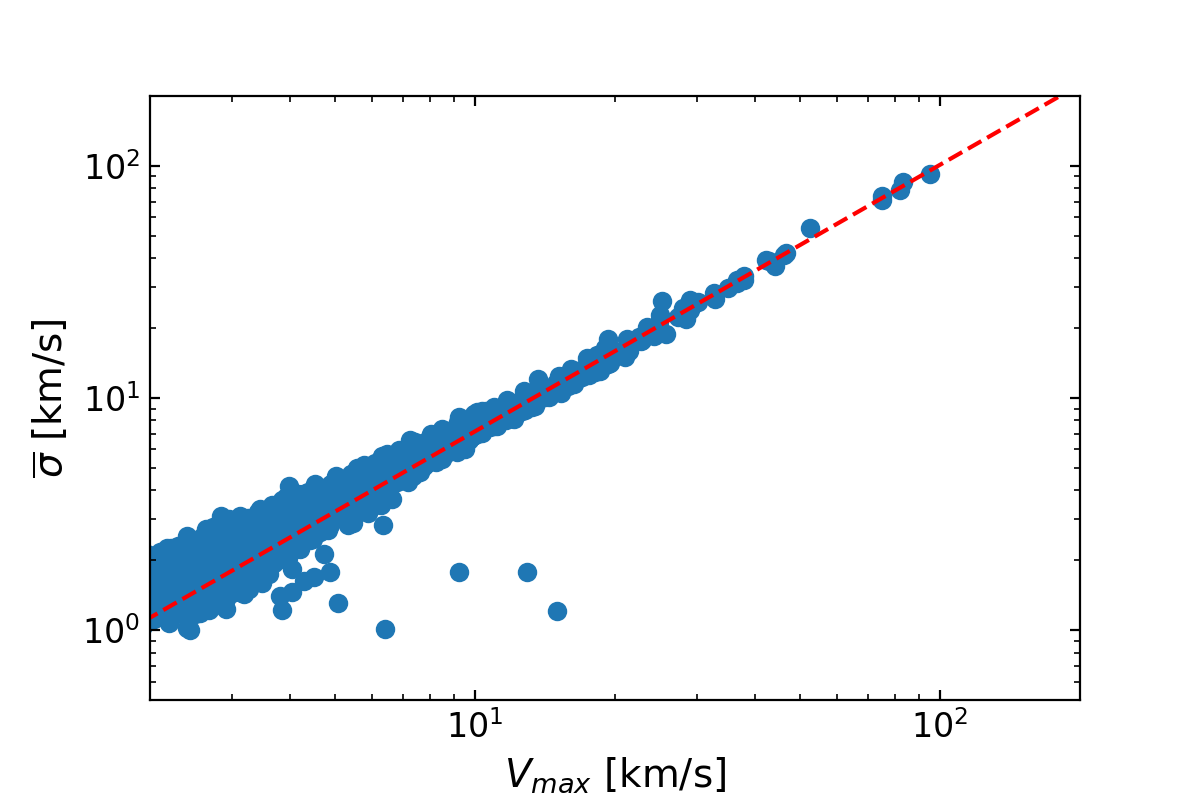}
    \caption{The distribution in $V_{\rm max}-\overline{\sigma}$ space of subhalos from all six Auriga halos within $R_{200}$ of each respective halo. We find $\overline{\sigma}$ by calculating the DM relative velocity distribution in radial shells, fitting a Maxwell-Boltzmann curve to the distribution with a 1D velocity dispersion $\sigma$, and then taking the average of these $\sigma$ values over all shells for each subhalo. The dashed red line indicates the best-fit power-law curve found to be $\overline{\sigma} = 0.51 (V_{\rm max})^{1.15}$, which we will use to extrapolate to lower mass subhalos. A similar result is obtained from the DMO simulations, where $\overline{\sigma} = 0.55 (V_{\rm max})^{1.10}$. Our results are consistent with the power law relation found in ref.~\cite{Blanchette:2022}.}
    \label{fig:maxwellsigma}
\end{figure}

\subsection{Relative velocity distributions}
\label{sec:relative velocities}
We now discuss the DM relative velocity distributions of subhalos using the notation established in our previous work~\cite{Board:2021bwj}. For each subhalo, we write the probability distribution of DM particles associated with only that subhalo as
\begin{equation}
P_{\mathbf{x}} (\mathbf{v}) = \frac{f(\mathbf{x}, \mathbf{v})}{\rho(\mathbf{x})},
\label{eq:P_x}
\end{equation}
where $\mathbf x$ is the position vector, $\mathbf v$ is the velocity vector, and the DM density at a position $\mathbf x$ in the subhalo is normalized as
\begin{equation} 
\rho ({\mathbf x}) = \int f ({\mathbf x},{\mathbf v}) \rm d^3 {\mathbf v}. 
\label{eq:rho} 
\end{equation}

The velocity vectors of the DM particles are determined with respect to the center of the main halo, whereas the position vectors are determined with respect to the center of the respective subhalo. This is appropriate as we are calculating the DM relative velocity distribution of particles within each subhalo, so that the bulk subhalo motion is subtracted out. Using spherical shells as defined in section~\ref{sec:Density profiles}, we resolve the velocity vectors into three components then subtract the components of the velocities in this basis, being careful to avoid double counting. We then take the modulus of the components of the pairwise relative velocities, which provides an estimate of $P_{\mathbf x} (\mathbf {v}_{\rm rel})$. In each radial shell, $P_{\mathbf x} (|\mathbf {v}_{\rm rel}|)$ is normalized to unity, such that
\begin{equation} 
\int P_{\mathbf x}(|\mathbf v_{\rm rel}|) ~d v_{\rm rel}= 1 
\end{equation} 
and therefore we have $\int P_{\mathbf x}(\mathbf v_{\rm rel}) ~d^3 \mathbf v_{\rm rel}= 1$.

\par Though there is some variation in the velocity distribution of subhalos, ref.~\cite{Blanchette:2022} shows that for MW dwarf spheroidal analogues, $P_{\mathbf x} (|\mathbf {v}_{\rm rel}|)$ can be well approximated by a Maxwell-Boltzmann (MB) distribution,
\begin{equation}
P_{\rm MB}(|\mathbf{v}_{\rm rel}|) = \sqrt{\frac{2}{\pi}}\frac{v_{\rm rel}^2}{\sigma^3}~\exp\left(-\frac{v_{\rm rel}^2}{2\sigma^2}\right),
\label{eq:MaxBdist}
\end{equation}
where $\sigma$ is the 1D relative velocity dispersion.

In the analysis of ref.~\cite{Blanchette:2022}, the best fit MB distribution is found in spherical shells at different radii from the center of the subhalo. Then the mean best fit peak speed of the MB distributions over all shells is calculated for each subhalo. Ref.~\cite{Blanchette:2022} finds that velocity-dependent $\mathcal{J}$-factors can be accurately estimated using the mean best fit MB parameters in simulated MW dwarf spheroidal galaxies. 

Following the same procedure as in ref.~\cite{Blanchette:2022}, we fit a MB distribution to the DM relative velocities in different spherical shells in each subhalo, and calculate the mean of the velocity dispersion, $\overline{\sigma}$, over all spherical shells in each subhalo. Figure~\ref{fig:maxwellsigma} shows the relationship between the $V_{\rm max}$ of a given resolved subhalo and the $\overline{\sigma}$ of the best-fit Maxwellian. The dashed line indicates the best-fit power-law curve which we will use to extrapolate to low-mass subhalos.

\section{Annihilation luminosities}
\label{sec:luminosities}

The annihilation luminosity from the DM particles is calculated from the DM density and the DM relative velocity distribution at each point within the halo. For the general case of velocity-dependent models, the annihilation luminosity from some region of space can be written as
\begin{equation}
L_{n}=\int {\rm d}^3 \mathbf x \int {\rm d}^3 \mathbf v_{\rm rel}
P_{\mathbf x} (\mathbf v_{\rm rel}) \left(\frac{{v}_{\rm rel}}{c}\right)^n
\left[\rho (x)\right]^2.
\end{equation}
For our velocity-dependent models, we examine the following possibilities: $n=-1$ (Sommerfeld-enhanced annihilation), $n=0$ (s-wave annihilation), $n=2$ (p-wave annihilation), and $n=4$ (d-wave annihilation). The different cross section models correspond to different velocity moments of the relative velocity distribution~\cite{Board:2021bwj},
\begin{equation} 
\mu_{n}(\mathbf x) \equiv \int {\rm d}^3 \mathbf v_{\rm rel} P_{\mathbf x} (\mathbf v_{\rm rel}) v_{\rm rel}^n,
\label{eq:velocity moment}
\end{equation}
where $\mu_{n}(\mathbf x)$ is the $n$-th moment of the relative velocity distribution $P_{\mathbf x} (\mathbf {v}_{\rm rel})$. In terms of the velocity moments, the annihilation luminosity can be written as
\begin{equation}
L_{n}=\int {\rm d}^3 \mathbf x \left[\rho (x)\right]^2 \left(\frac{\mu_{n}(\mathbf x)}{c^n}\right).
\label{eq:L_n}
\end{equation}

The annihilation luminosity has contributions from both the smooth halo and the subhalo components. We start by estimating the annihilation luminosity from the smooth component of the DM halo. Using the Voronoi tessellation method described above, we estimate the local DM density at the location of each DM particle. Then we calculate the relative velocity distribution at each point on a spherical grid, using the nearest 500 DM particles. The relative velocity distributions are then used in eq.~\eqref{eq:velocity moment} to obtain $\mu_{n}(\mathbf x)$ for each annihilation model at each point. We interpolate these results to obtain the relative velocity moments at the location of each DM particle in the smooth halo. We then compute the integral in eq.~\eqref{eq:L_n} over each volume produced by the Voronoi Tessellation to obtain the annihilation luminosity produced by each DM particle in the smooth halo.

\par For the subhalo component, we calculate the annihilation luminosity by splitting up the contribution from large and small subhalos. As mentioned in section~\ref{sec:Density profiles}, we consider large subhalos to have an $R_{\rm max}$ which corresponds to an angular size $>1$ degree as seen from the solar position of $8.0$~kpc, whereas we consider small subhalos to have an $R_{\rm max}$ which corresponds to an angular size of $<1$ degree. We use this angular size definition for large and small subhalos, because $1$ degree corresponds to the approximate angular resolution scale for Fermi-LAT at the energies relevant for DM searches. For large subhalos in the simulations, we calculate the annihilation luminosity from each DM particle using methods similar to that of the smooth halo component. The only difference is that the relative velocity moments, $\mu_{n}(\mathbf x)$, are calculated using a Maxwell-Boltzmann distribution with a dispersion set equal to the mean velocity dispersion, $\overline{\sigma}$, computed as described in section~\ref{sec:relative velocities}. We use the mean dispersion to estimate the relative velocity moment for all points within the subhalo, and therefore $\mu_{n}$ would be independent of the position vector in the subhalos.

\begin{figure}[t]
    \centering
    \includegraphics[width=1.0\textwidth]{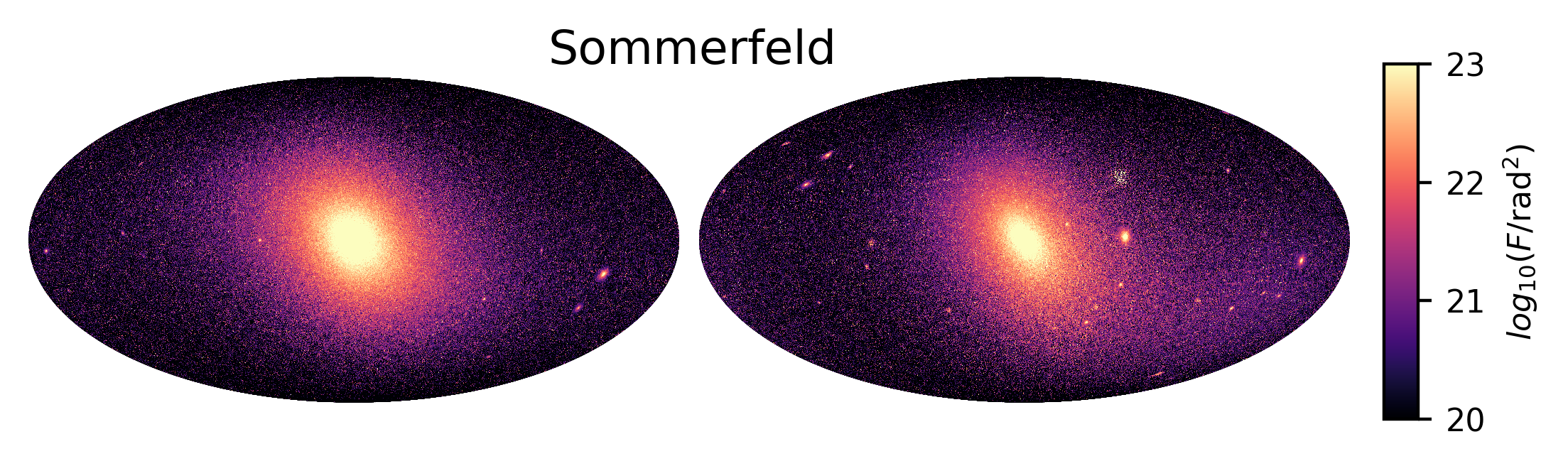}
    \includegraphics[width=1.0\textwidth]{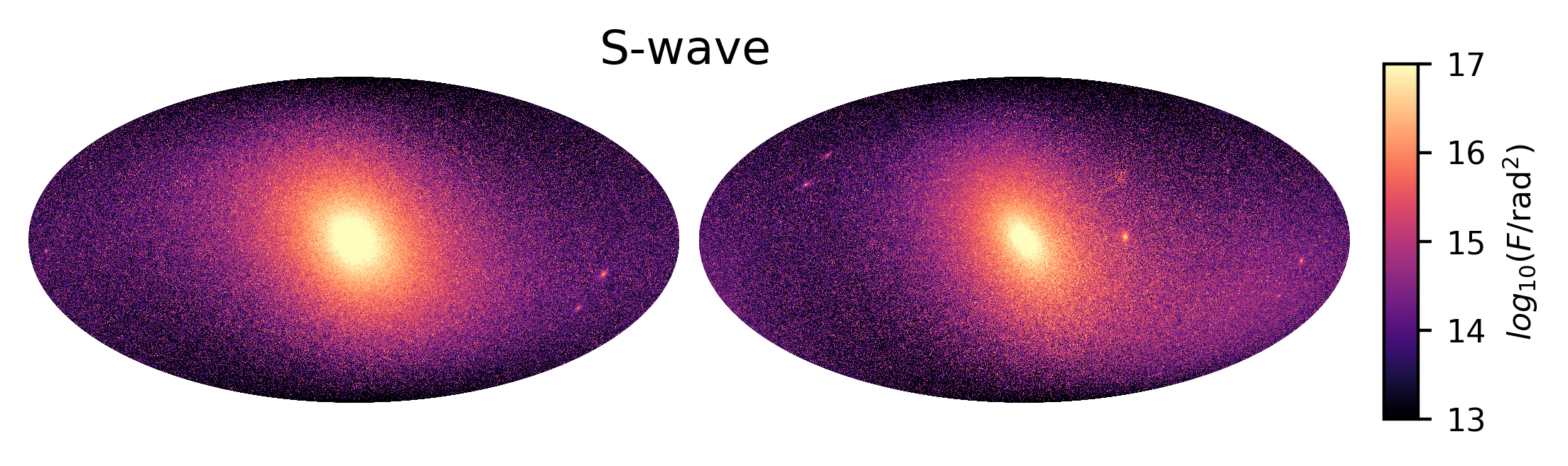}
    \includegraphics[width=1.0\textwidth]{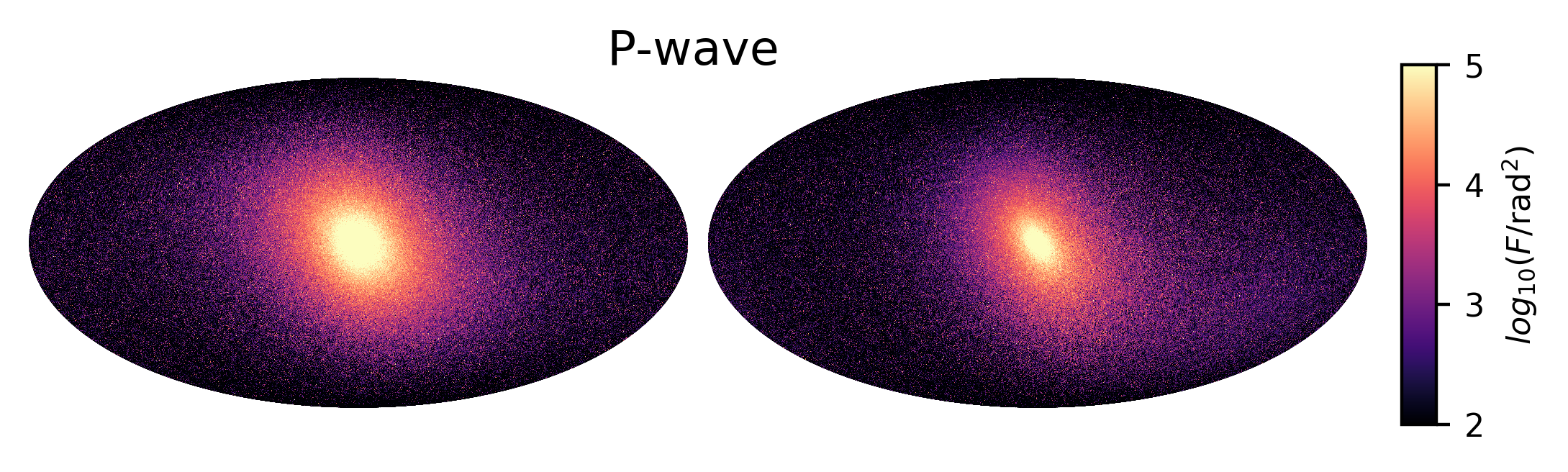}
    \includegraphics[width=1.0\textwidth]{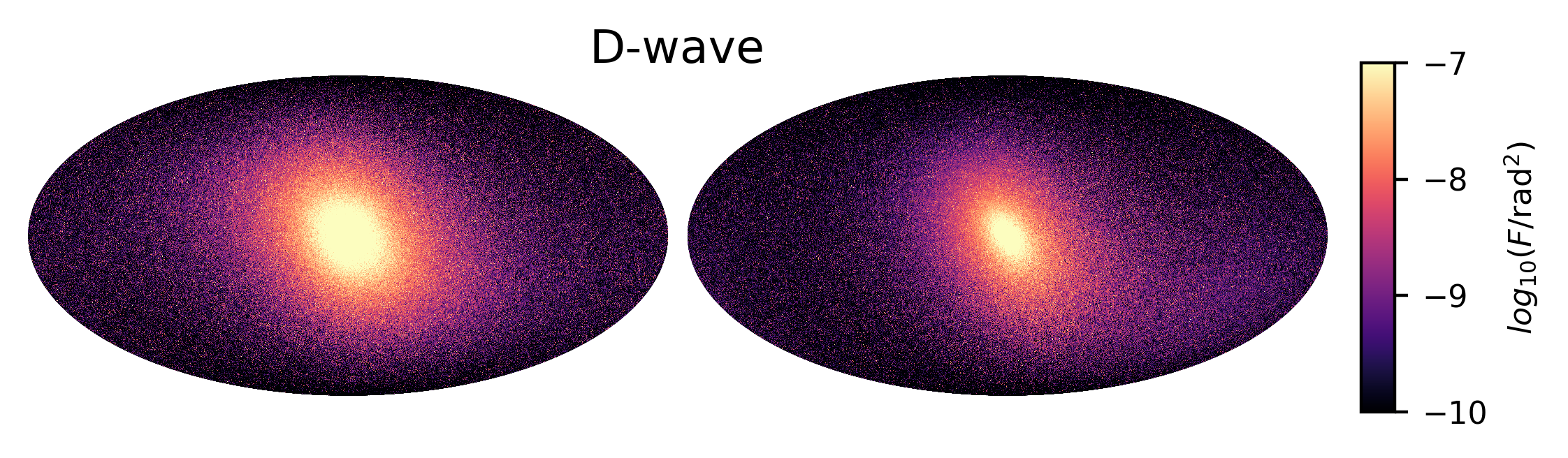}
    \caption{All-sky Mollweide projections of the DM annihilation flux density for each considered annihilation model as seen from the midplane of the stellar disc, 8.0 kpc from the Galactic center of the Au6 halo (left) and its DMO counterpart (right). The color bars to the right of each pair of sky maps show the approximate range of the annihilation flux density for each annihilation model. The DM annihilation fluxes from subhalos are clearly systematically fainter compared to the smooth halo component in all annihilation models than in the Sommerfeld model, with the faintest subhalos shown in the d-wave model.}
    \label{fig:skymap}
\end{figure}

For smaller subhalos in our simulations, whose $R_{\rm max}$ has an angular size less than $1.0$ degree, we estimate the total DM annihilation luminosity from a spherical region interior to $R_{\rm max}$ as
\begin{equation}
L_{\rm sub}=\frac{C_{\rm Einasto}V_{\rm max}^4}{G^2R_{\rm max}},
\label{eq:Lsub}
\end{equation}
where $G$ is the gravitational constant and $C_{\rm Einasto} = 1.87$ for an Einasto density profile with $\alpha = 0.16$~\cite{Grand:2020}. Since we have chosen a $\mu_{n}$ that is not dependent on position for subhalos, we can then rewrite eq.~\eqref{eq:L_n} as
\begin{equation}
\begin{split}
L_{n,{\rm sub}}&=\left(\frac{\mu_{n}}{c^n}\right)\int {\rm d}^3 \mathbf x \left[\rho (x)\right]^2 \\
&= \left(\frac{\mu_{n}}{c^n}\right)L_{\rm sub}\\
&= \left(\frac{\mu_{n}}{c^n}\right)\left(\frac{C_{\rm Einasto}V_{\rm max}^4}{G^2R_{\rm max}}\right) .
\label{eq:L_n subhalos}
\end{split}
\end{equation}

Then including the contribution from the smooth component and the subhalos, we examine these luminosities from one solar position at 8.0 kpc from the Galactic center by calculating the annihilation flux as in ref.~\cite{Grand:2020},
\begin{equation}
F=L/d^2,
\end{equation}
where $L$ is the luminosity of a subhalo or DM particle and $d$ is the heliocentric distance of that subhalo or DM particle. We sum the annihilation flux from the smooth DM halo, large DM subhalos, and small DM subhalos in bins of equal angular size of $1.9\times10^{-5}$ rad$^2$. The results of the flux density for each annihilation model are shown in figure~\ref{fig:skymap}. The all-sky Mollweide projection maps on the left are the results for Au6 and those on the right are for its DMO counterpart. For each annihilation model, we find that the smooth component of the DM halo is brighter and rounder in shape in the hydrodynamical simulations compared to their DMO counterparts. We also find that subhalo fluxes are systematically fainter in the hydrodynamical simulations than their DMO counterparts for each annihilation model, consistent with previous results that examined s-wave models~\cite{Grand:2020}. When comparing the subhalo fluxes for different annihilation models in the same simulation, we find that the subhalo flux relative to the flux from the smooth halo component appears to be largest for the Sommerfeld model. Subhalo fluxes are suppressed relative to the smooth halo in the p-wave and d-wave models, which we will quantify in section~\ref{sec:results}.

\section{Results}
\label{sec:results}
\par In this section we present the primary results of our analysis. We begin by comparing the contribution to the luminosity from subhalos and the smooth halos in the simulations. We then characterize the contribution of the integrated subhalo luminosity due to subhalos across different luminosity scales. We also estimate the impact of extrapolating the luminosity function of subhalos below the lowest mass ($\sim 10^6~\Msun$) subhalos resolved in the simulations.

\begin{figure}[t]
    \centering
    \includegraphics[width=0.49\textwidth]{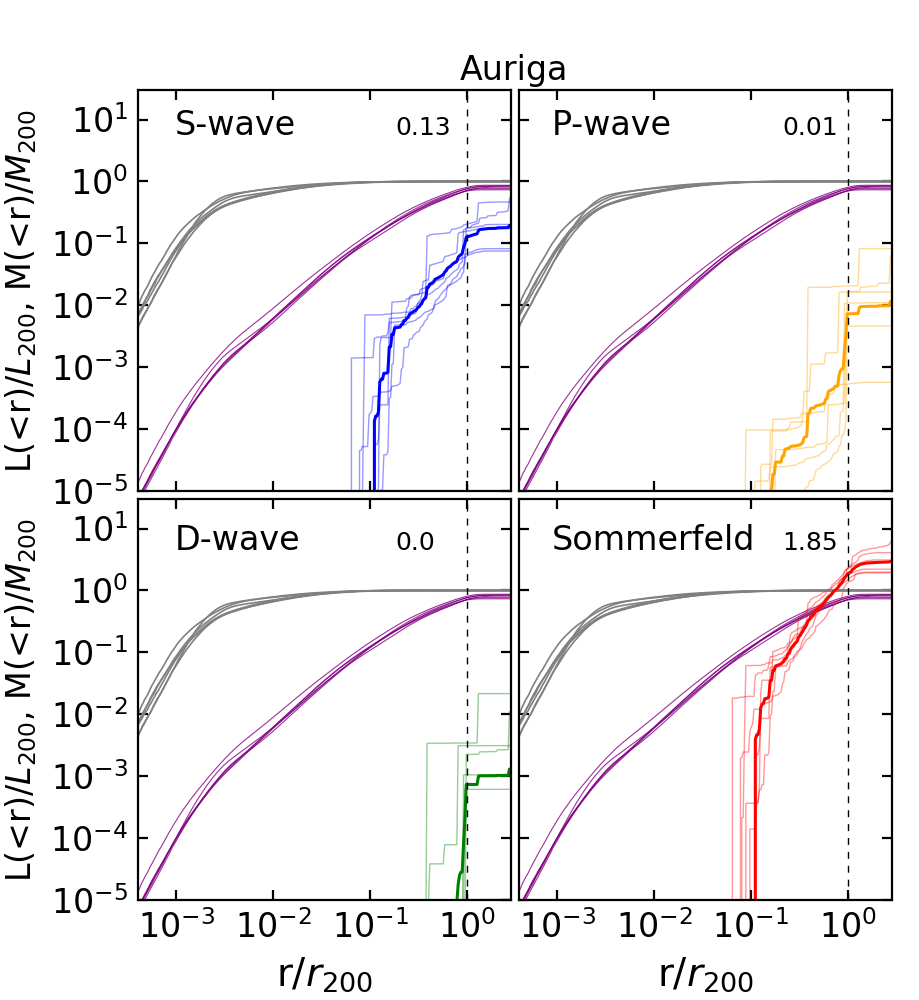}
     \includegraphics[width=0.49\textwidth]{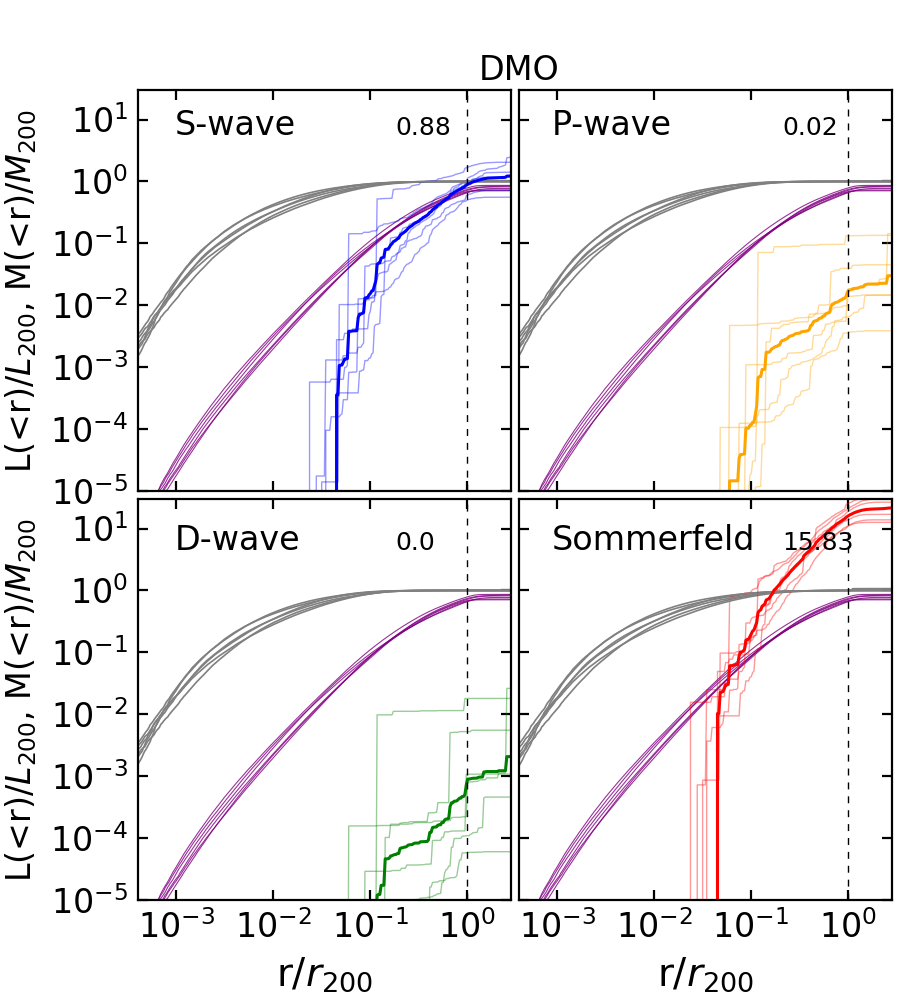}
    \caption{The total DM annihilation luminosity within some Galactocentric radius, $r$, as a function of that radius for the Auriga halos (left four panels) and their DMO counterparts (right four panels). The gray lines are the results from the smooth DM halo component of the six Auriga halos. The results for the resolved DM subhalos in each halo are shown as thin blue (s-wave), yellow (p-wave), green (d-wave), and red (Sommerfeld) lines. The thick lines of the same color correspond to the average total luminosity of resolved subhalos across all six Auriga halos. In each panel, the number in the upper right indicates the average total luminosity from subhalos within $r_{200}$. Both luminosities from the smooth halo component and the subhalos have been normalized by the total luminosity, $L_{200}$, within $r_{200}$ for the corresponding smooth halo component for each annihilation model. The purple lines indicate the total smooth mass within $r$ for each halo, normalized by the total mass, $M_{200}$, within $r_{200}$. The dashed vertical lines indicate $r_{200}$ for all halos. The luminosities from the velocity-independent s-wave annihilation model agree with the results of ref.~\cite{Grand:2020}.}
    \label{fig:totalluminosityhydro}
\end{figure}

\subsection{Luminosities of the smooth halos and resolved subhalos}

Figure~\ref{fig:totalluminosityhydro} shows the DM annihilation luminosities of six Auriga halos within a distance $r/r_{200}$, where $r$ is the radial distance from the Galactic center and $r_{200}$ is the virial radius of each halo. Shown are both the contributions from the smooth DM halo and from the subhalos for each of the six Auriga halos (left four panels) and the results for their DMO counterparts (right four panels). The gray lines indicate the luminosity from the smooth DM halo of each simulated Auriga halo for each annihilation model, while the thin blue, yellow, green, and red lines show the luminosity from all subhalos within an Auriga halo for the s-wave, p-wave, d-wave, and Sommerfeld models, respectively. For each model, the thick lines of the same color show the average total subhalo luminosity across all six halos. The purple lines show the total mass of the smooth component within radius $r$ for each halo.

We compare the results for the average total subhalo luminosity to the total smooth halo luminosity. For the Auriga halos, we find that the luminosity from the smooth DM halo dominates over the average luminosity from subhalos in all annihilation models except for the Sommerfeld model, where the subhalo luminosity dominates at $r/r_{200} > 0.74$. For the DMO halos, we find that the luminosity from the smooth DM halo dominates for p-wave and d-wave annihilation, but the average luminosity from subhalos surpasses that of the smooth DM component at $r/r_{200} > 1.2$ for the s-wave model and at $r/r_{200} > 0.17$ for Sommerfeld.

Next we compare the results from the Auriga halos to that of their DMO counterparts. Examining the smooth halo components, we see that the luminosities of the Auriga halos approach $L_{200}$ more rapidly at smaller radii than their DMO counterparts, which is also illustrated by the brighter central regions in the sky maps in figure~\ref{fig:skymap}. This effect is a result of the contraction of the central regions of the smooth DM halos due to the presence of baryons. For a given annihilation model, we find that the subhalo luminosities in the DMO simulations typically have a larger value at the same distance $r/r_{200}$. This is due to baryonic processes in which the baryonic disc preferentially destroys nearby subhalos.

\subsection{Subhalo luminosity functions} 

\begin{figure}[t]
    \centering
    \includegraphics[width=0.98\textwidth]{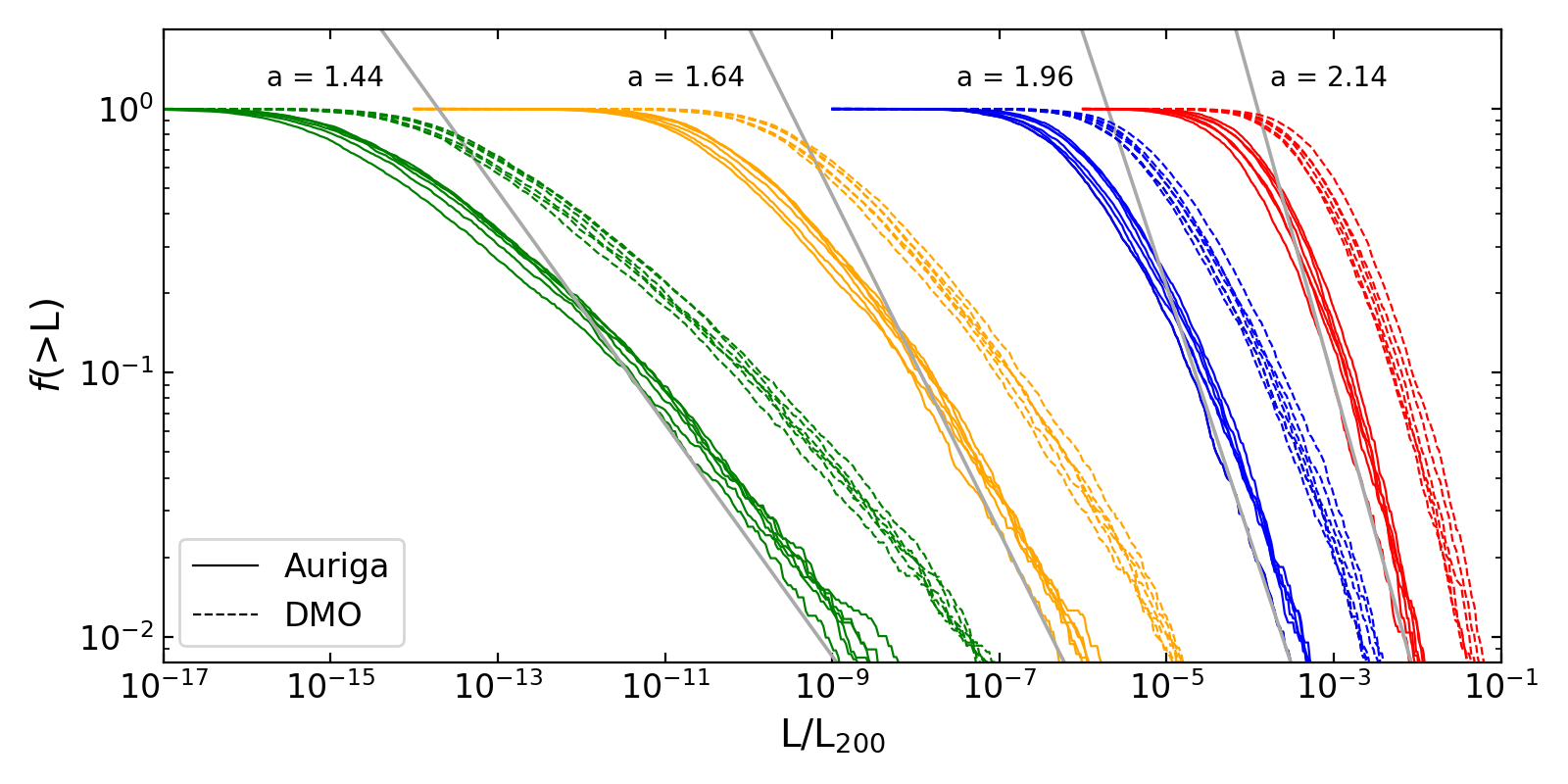}
    \caption{The fraction of subhalos with luminosity greater than some luminosity, $L$, relative to the total luminosity, $L_{200}$, within $r_{200}$ of each smooth halo component. For s-wave (blue), p-wave (yellow), d-wave (green), and Sommerfeld (red) annihilation models we show the results for both the Auriga halos and their DMO counterparts in solid and dashed lines, respectively. The gray lines indicate the best fit power law for Au6 over the range of luminosities associated with subhalos with $20~{\rm km/s} \leq V_{\rm max} \leq 60$~km/s. The best fit values of $a$ from eq.~\eqref{eq: luminosity fraction} for Au6 are listed next to the corresponding gray line.}
    \label{fig:luminosityfraction}
\end{figure}

We now move on to analyze the subhalo differential luminosity functions, $dN/dL$, where $N$ is the number of subhalos with luminosity $L$, for each of our annihilation models. From this definition of the differential luminosity function, we construct the fraction of subhalos with luminosity greater than $L/L_{200}$ within each simulation for each DM annihilation model, 
\begin{equation}
f(>L) \equiv \frac{\int_L^{L_{max}} L' \frac{dN}{dL'}dL'}{\int_{L_{min}}^{L_{max}} L' \frac{dN}{dL'}dL'}.
\end{equation}

\par To provide a physical interpretation for the cumulative luminosity function, we compare to a power law defined as 
\begin{equation}
f(>L) \propto L^{-(a-1)}.
\label{eq: luminosity fraction}
\end{equation}
Defined in this manner, the subhalo luminosity function is dominated by the highest (lowest) luminosity subhalos for $a < 2$ ($a > 2$). In order to conservatively avoid the impact of numerical resolution, we calculate this quantity over the range of $L$ calculated from subhalos with $20~{\rm km/s} \leq V_{\rm max} \leq 60$ km/s for each simulation. 

\par Figure~\ref{fig:luminosityfraction} shows the cumulative luminosity function, $f(>L)$, within each simulated halo for each DM annihilation model. S-wave (blue), p-wave (yellow), d-wave (green), and Sommerfeld (red) annihilation models are shown for the Auriga halos (solid lines) and their DMO counterparts (dashed lines). The gray lines and corresponding $a$ values indicate the best fit parameters of eq.~\eqref{eq: luminosity fraction} for Au6. For the Auriga halos we find the range of $a$ values to be $[1.80-2.07]$ for s-wave, $[1.45-1.66]$ for p-wave, $[1.38-1.49]$ for d-wave, and $[2.10-2.43]$ for the Sommerfeld model. For the DMO counterparts we find the range of $a$ values to be $[1.75-1.99]$ for s-wave, $[1.50-1.61]$ for p-wave, $[1.35-1.44]$ for d-wave, and $[ 1.94-2.34]$ for the Sommerfeld model. 

\par These fit results indicate that for the case of the Sommerfeld model, the integrated subhalo luminosity is dominated by the least luminous subhalos, while for s, p and d-wave models, the luminosity is dominated by the most luminous subhalos. Going from s to d to p-wave, the high luminosity subhalos become more and more significant as a fraction of the total subhalo emission, even though similarly going from s to d to p-wave, the total luminosity contribution from subhalos becomes progressively smaller as compared to the smooth halo.

 \begin{figure}[t]
    \centering
    \includegraphics[width=0.49\textwidth]{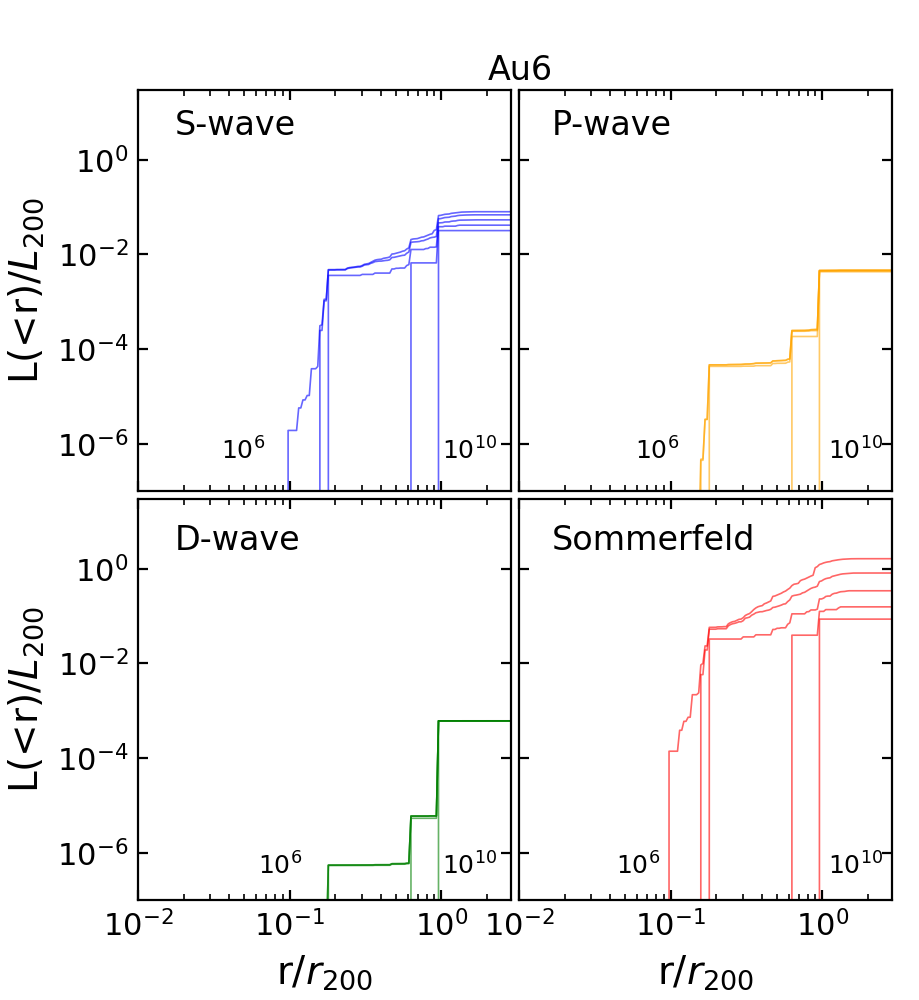}
    \includegraphics[width=0.49\textwidth]{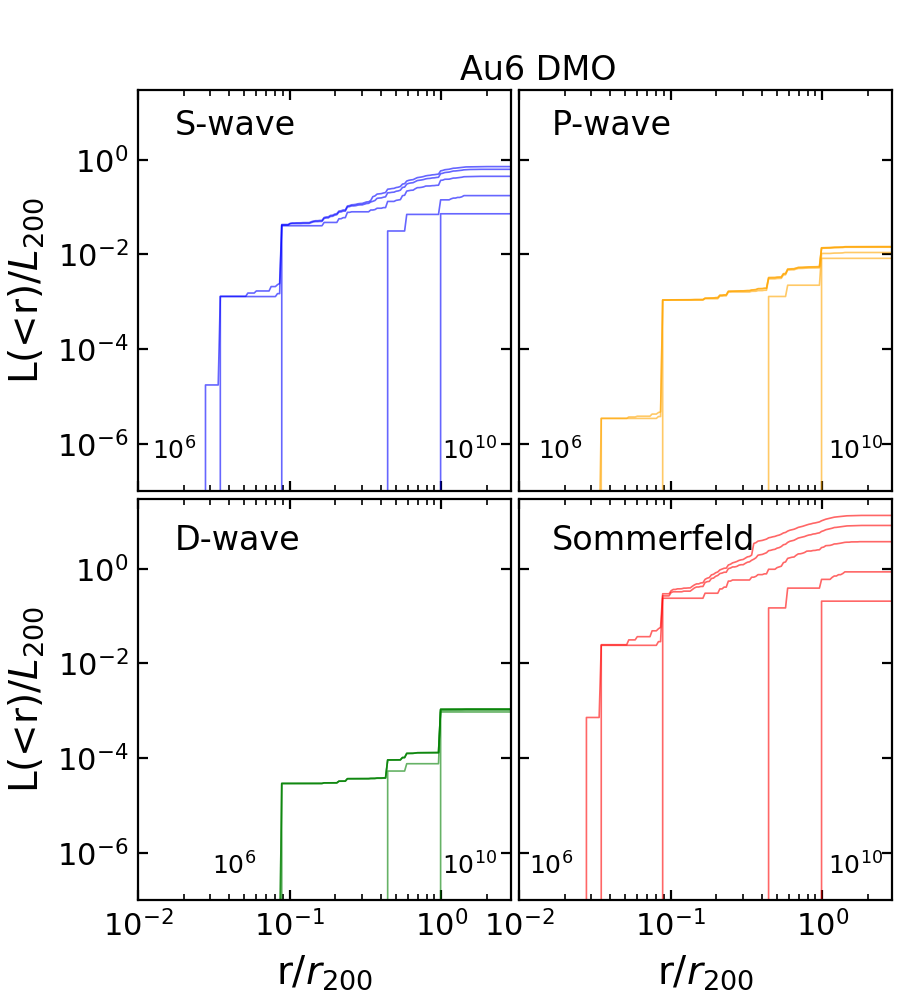}
    \caption{The luminosity contribution from Au6 subhalos for each annihilation model for various lower limits of the subhalo DM mass. Each line corresponds to a minimum DM mass of $10^{6}, 10^{7}, 10^{8}, 10^{9},$ or $10^{10}~\Msun$. The lines corresponding to higher mass subhalos trend to the right side of each panel. This shows that most of the total subhalo luminosity is due to large subhalos in the s-wave, p-wave, and d-wave annihilation models. However, in the case of Sommerfeld annihilation there is a notable fraction of luminosity from low-mass subhalos. We find similar results for all six halos.}
    \label{fig:subhaloluminosityhydro}
\end{figure}

 \par To further examine the contributions to the luminosity from different subhalo mass intervals, the four panels on the left side of Figure~\ref{fig:subhaloluminosityhydro} show the contribution to the subhalo luminosity from Au6 for different DM mass ranges and the four panels on the right side show the results for the DMO counterpart. We consider the luminosities of all subhalos with DM masses above a minimum mass of $10^{6}, 10^{7}, 10^{8}, 10^{9},$ and $10^{10}~\Msun$. We calculate the total luminosity from subhalos within $r_{200}$ and above a minimum mass of $10^{8}~\Msun$ as a fraction of the total luminosity from resolved subhalos within $r_{200}$. For Au6 we find the luminosity fraction to be 0.689 for s-wave, 0.996 for p-wave, 1.000 for d-wave, and 0.177 for Sommerfeld. For the DMO counterpart we find the luminosity fraction to be 0.626 for s-wave, 0.977 for p-wave, 0.999 for d-wave, and 0.264 for Sommerfeld. For s-wave, p-wave, and d-wave annihilation models, a large part of the total luminosity is due to high-mass subhalos. Whereas in the case of the Sommerfeld model we find that there is a large contribution to the total subhalo luminosity from low-mass subhalos.

\subsection{Low-mass subhalo extrapolation}
As discussed above, the Auriga simulations resolve DM subhalo masses down to $\sim 10^{6}~\Msun$. However, this is plausibly still much larger than the cut-off mass in cold DM, which may be as small as Earth mass~\cite{Frenk:2012}. It is interesting to estimate the effects that an extrapolation down to mass scales below the Aurgia resolution would have on our results. 

\par In this analysis, for computational convenience we extrapolate DM subhalos down to $\sim 10^{0}~\Msun$. To estimate the abundance of these low-mass subhalos below the resolution scale, we follow the work of ref.~\cite{Grand:2020}. For each halo, we estimate the overall abundance of subhalos in the range $0.1~{\rm km/s} \leq V_{\rm max} \leq 10$~km/s, with a subhalo of maximum circular velocity $0.1$~km$/$s corresponding approximately to a subhalo of mass $10^{0}$ M$_{\odot}$. We assign a $V_{\rm max}$ value to each of the extrapolated subhalos using the differential $V_{\rm max}$ function in ref.~\cite{Grand:2020}. The subhalo is then assigned an $R_{\rm max}$ value using the median $R_{\rm max}-V_{\rm max}$ relation shown in ref.~\cite{Grand:2020} and derived from ref.~\cite{Wang:2020} for extrapolation to lower subhalo masses. These values for $V_{\rm max}$ and $R_{\rm max}$ are then used in eq.~\eqref{eq:Lsub} to estimate the velocity-independent s-wave annihilation luminosity for extrapolated subhalos.

Figure~\ref{fig:maxwellsigma} above shows the relationship between the $V_{\rm max}$ of a given resolved subhalo and the $\overline{\sigma}$ value associated with that subhalo as discussed in section~\ref{sec:relative velocities}. The dashed line indicates the best-fit power-law curve which we will use to extrapolate to low-mass subhalos. The extrapolated $\overline{\sigma}$ values are used to produce a Maxwellian distribution, which is then used in eq.~\eqref{eq:velocity moment} as the probability distribution, $P_{\mathbf x}$, to calculate the velocity moment, $\mu_{n}$, for each of the velocity-dependent annihilation models. The extrapolated velocity moments, along with the $V_{\rm max}$ and $R_{\rm max}$ calculated above, are then used in eq.~\eqref{eq:L_n} to estimate the velocity-dependent annihilation luminosity for extrapolated subhalos.

Given the structural properties of the extrapolated subhalos, we then must assign them a position within the halo. To assign the position, we start from a spherically symmetric number density distribution that is generated from the resolved subhalos with $10~{\rm km/s}< V_{\rm max} <30$~km/s. We then fit a power-law curve to this number density profile, which we use to radially distribute the subhalos. These objects are then distributed randomly in the angular coordinates, which produces a spherically symmetric distribution of subhalos.

\begin{figure}[t]
    \centering
    \includegraphics[width=0.49\textwidth]{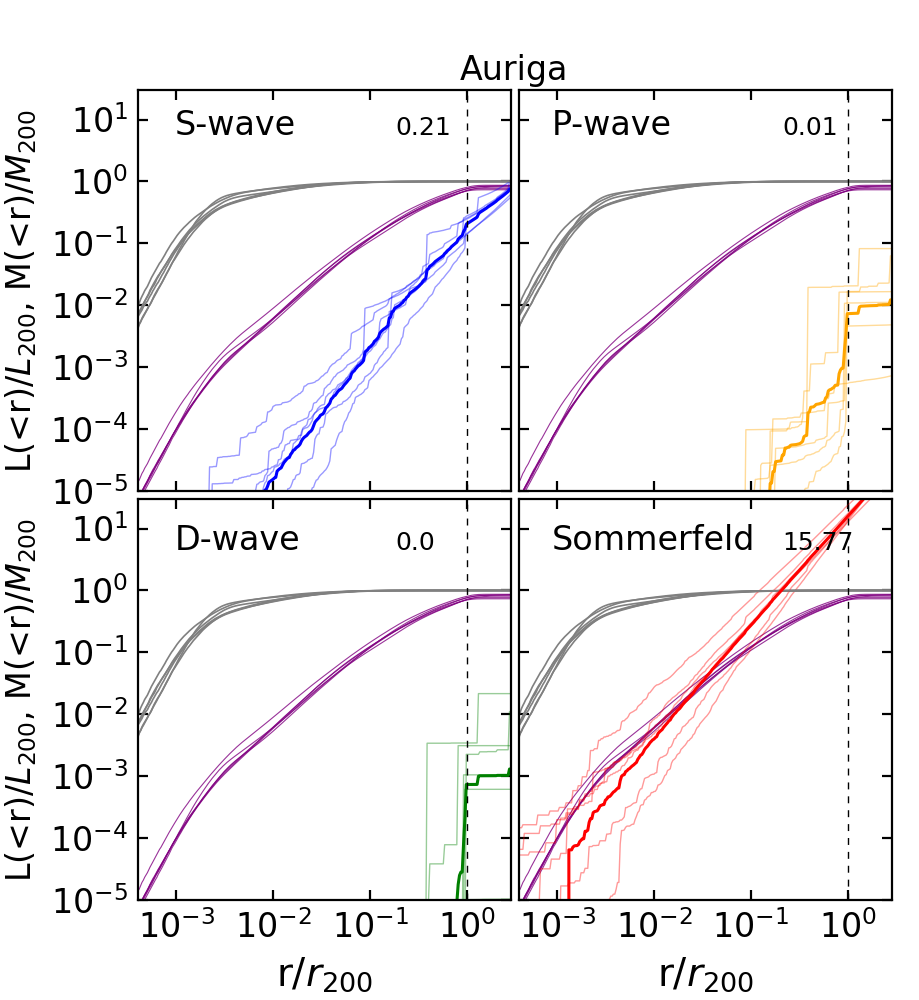}
    \includegraphics[width=0.49\textwidth]{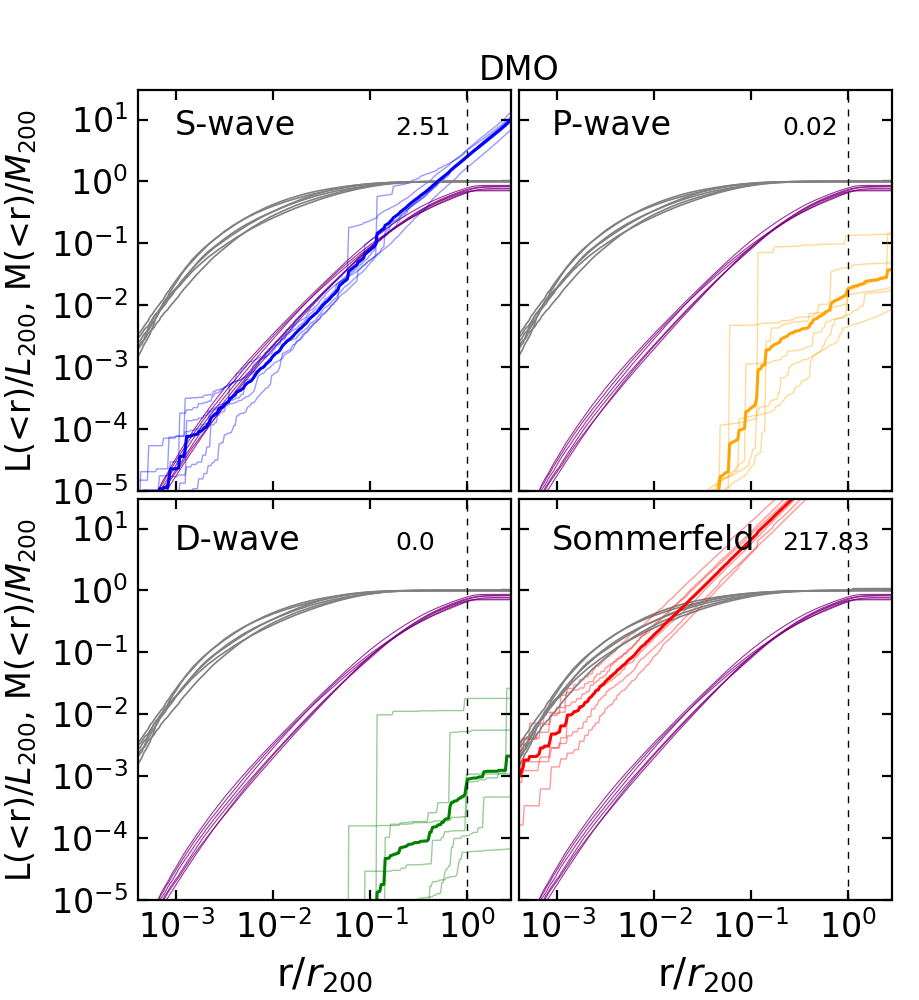}
    \caption{Same as figure~\ref{fig:totalluminosityhydro}, but including the results for extrapolated subhalos with $0.1~{\rm km/s} \leq V_{\rm max} \leq 10$~km/s. In each panel, the number in the upper right indicates the average total luminosity due to resolved and extrapolated subhalos within $r_{200}$, normalized by $L_{200}$.}
    \label{fig:extrapolatedluminosityhydro}
\end{figure}

The effects of the addition of extrapolated subhalos on the total luminosity for all six Auriga halos are depicted in the left four panels of figure~\ref{fig:extrapolatedluminosityhydro} and the results for their DMO counterparts are shown in the right four panels. For the Auriga halos, we still find that the luminosity from the smooth DM halo dominates over the average luminosity from subhalos in all models except for Sommerfeld annihilation, which now dominates at $r/r_{200} > 0.2$ rather than at $r/r_{200} > 0.74$ in the case of only resolved subhalos. For the DMO counterparts, we find that the average luminosity from subhalos now dominates for the s-wave model at $r/r_{200} > 0.5$ and for the Sommerfeld model at $r/r_{200} > 0.03$. 

\section{Conclusions and discussion}
\label{sec:conclusion}

\par We have used the Auriga simulations of Milky Way-like galaxies to determine the contribution of halo substructure to the signal from dark matter annihilation. We consider the general case of velocity-dependent dark matter annihilation, examining Sommerfeld, s-wave, p-wave, and d-wave models. We find that substructure is the most significant in Sommerfeld models, while it is the least significant in d-wave models. In the Sommerfeld models, the substructure contribution to the dark matter annihilation signal dominates that of the smooth component beyond $\sim 0.74 r_{200}$, while for all other models the substructure contribution is sub-dominant at all radii as compared to the smooth halo. 

\par Examining the luminosity functions of substructure, we find that in Sommerfeld models, the luminosity function is dominated by the least luminous subhalos that are resolved. On the other hand, for d-wave models, the luminosity function is dominated by the most luminous subhalos that are resolved. So extrapolating to lower subhalo mass scales may still increase the luminosity contribution from subhalos in Sommerfeld models, though it will not affect the luminosity contribution from subhalos in the case of d-wave models. 

Systematic uncertainties in the results can also arise from the uncertainty in the number of subhalos, due to a different treatment of baryonic effects. We do not expect these uncertainties  to affect our results for the d-wave model, since the luminosities are not centrally concentrated in that case. However, they can introduce additional uncertainties in our results for the Sommerfeld model. Studying the $\mathcal{J}$-factors in an even larger sample of simulations is  important to quantify such uncertainties.

Another source of systematic uncertainty can arise from the assumed DM density profile of the subhalos. In our analysis, we have assumed the shape of the DM density profile interior to $R_{\rm max}$ to be an Einasto density profile, and we have estimated the total luminosity of each subhalo using $C_{\rm Einasto} = 1.87$ in eq.~\ref{eq:L_n subhalos}. Since $V_{\rm max}$, $R_{\rm max}$, and $\mu_n$ are calculated directly from the simulation data, the total luminosity may be  sensitive to the choice of DM density profile. To examine the systematic arising from the assumed profile, we consider simply how our results change when assuming an NFW profile instead of an Einasto profile. For the NFW case, we can estimate the total luminosity from a subhalo using an NFW profile with $C_{\rm NFW} = 1.23$~\cite{Springel:2008zz}. This implies that using an NFW profile instead of an Einasto profile would simply scale our results by $\gtrsim 30\%$,  $L_{\rm sub}^{\rm NFW} = 0.66\,L_{\rm sub}^{\rm Einasto}$, for all annihilation models.

\par The results of our analysis have interesting implications for gamma-ray emission observed by Fermi-LAT. For example, the Galactic Center Excess (GCE) does not yet have a clear explanation, and may be consistent with particle dark matter annihilation~\cite{Hooper:2010mq}. However, it is possible that this emission is inconsistent with limits obtained from dwarf galaxies~\cite{Murgia:2020dzu}. Including the full effect of baryonic physics, the morphology of the GCE~\cite{DiMauro:2021} is consistent with the signal from the smooth component of the dark matter distribution~\cite{Grand:2022olu}. Since the morphology of the smooth emission component is similar for Sommerfeld, s-wave, p-wave, and d-wave models, the GCE would similarly be well fit by the smooth component of any of these velocity-dependent models. The bounds on the cross section would simply scale with the ratio of the $\mathcal{J}$-factors, in a manner similar to that discussed in Ref.~\cite{McKeown_2022}. However, one caveat to this statement is that the simulations that we have considered resolve subhalos down to mass scales of $\gtrsim 10^6$ M$_\odot$. This may be far larger than the actual minimum subhalo mass, and an extrapolation down to lower subhalo masses may be particularly important for Sommerfeld models, in which case the subhalo component may eventually dominate over the smooth halo emission.

\par Another galaxy that our results may be considered in the context of is M31. Fermi-LAT has previously detected emission from the central regions of M31, which may be explained via cosmic-ray interactions in the central stellar disk~\cite{Fermi-LAT:2010zba}. More recently, there has been an indication of an extended emission from the region surrounding M31, which may be explained by emission from its more extended dark matter halo~\cite{Karwin:2020tjw}. The M31 system is a unique target for dark matter annihilation, because halo substructure is expected to contribute to the emission in the outermost regions. Our results indicate that, even in the context of the full physics simulations, substructure emission is significant for Sommerfeld models, and even in the case of s-wave models the total emission from subhalos nears that of the smooth component around $r_{200}$. However, for p and d-wave models, the smooth component is dominant at all radii, and no emission from substructure would be identified. This shows that M31 provides a unique system for dark matter annihilation and substructure analysis, and we defer its detailed study to future work. 


\acknowledgments
LES and EP acknowledge support from DOE Grant de-sc0010813. This work was supported by a Development Fellowship from the Texas A$\&$M University System National Laboratories Office. KB and NB acknowledge the support of the Natural Sciences and Engineering Research Council of Canada (NSERC), funding reference number RGPIN-2020-07138. CSF acknowledges European Research Council (ERC) Advanced Investigator 
grant DMIDAS (GA 786910). RG acknowledges financial support from the Spanish Ministry of Science and Innovation (MICINN) through the Spanish State Research Agency, under the Severo Ochoa Program 2020-2023 (CEX2019-000920-S). 
This work used the DiRAC@Durham facility managed 
by the Institute for Computational Cosmology on behalf of the STFC DiRAC 
HPC Facility (www.dirac.ac.uk). The equipment was funded by BEIS capital 
funding via STFC capital grants ST/K00042X/1, ST/P002293/1, ST/R002371/1 
and ST/S002502/1, Durham University and STFC operations grant 
ST/R000832/1.  DiRAC is part of the National e-Infrastructure.

\bibliographystyle{JHEP}
\bibliography{./refs}

\end{document}